\documentclass[sigconf, twocolumn]{acmart}
\acmConference[EASE 2025]{The 29th International Conference on Evaluation and Assessment in Software Engineering}{17–20 June, 2025}{Istanbul, Turkey}

\AtBeginDocument{%
  }

\usepackage{amsmath}
\usepackage{booktabs}
\usepackage{multirow}
\usepackage{graphicx}
\usepackage{adjustbox}
\usepackage{fancyhdr}
\usepackage{listings}
\usepackage{graphicx}
\usepackage{tikz}
\newcommand{\vDotted}[1]{
  \tikz[baseline]{\draw[dotted, line width=0.6pt] (0,0) -- (0,#1);}%
}
\usepackage{xcolor}
\usepackage{balance}
\usepackage{enumitem}
\usepackage{subcaption}
\definecolor{skyblue}{rgb}{0.53,0.81,0.92}
\definecolor{lightorange}{RGB}{255, 200, 130}
\definecolor{grey}{gray}{0.5}

\begin{document}
\title{Co-Change Graph Entropy: A New Process Metric for Defect Prediction}
%


\author{Ethari Hrishikesh}
\affiliation{%
  \institution{Indian Institute of Information Technology, Manipur }
  \city{Mantripukhri, Imphal}
  \country{India}}
\email{hrishikeshethari@gmail.com}

\author{Amit Kumar}
\affiliation{%
  \institution{Indian Institute of Information Technology Allahabad}
  \city{Prayagraj}
  \country{India}
}
\email{amitchandramunityagi@gmail.com}

\author{Meher Bhardwaj}
\affiliation{%
 \institution{Indian Institute of Information Technology, Manipur}
 \city{Mantripukhri, Imphal}
 \country{India}}
 \email{bhardwajmeher01@gmail.com}

\author{Sonali Agarwal}
\affiliation{%
  \institution{Indian Institute of Information Technology Allahabad}
  \city{Prayagraj}
  \country{India}}
  \email{sonali@iiita.ac.in}


\begin{abstract}
 Process metrics, valued for their language independence and ease of collection, have been shown to outperform product metrics in defect prediction. Among these, change entropy (Hassan, 2009) is widely used at the file level and has proven highly effective. Additionally, past research suggests that co-change patterns provide valuable insights into software quality. Building on these findings, we introduce Co-Change Graph Entropy, a novel metric that models co-changes as a graph to quantify co-change scattering.
\par Experiments on eight Apache projects reveal a significant correlation between co-change entropy and defect counts at the file level, with a Pearson correlation coefficient of up to 0.54. In file-level defect classification, replacing change entropy with co-change entropy improves AUROC in 72.5\% of cases and MCC in 62.5\% across 40 experimental settings (five machine learning classifiers and eight projects), though these improvements are not statistically significant. However, when co-change entropy is combined with change entropy, AUROC improves in 82.5\% of cases and MCC in 65\%, with statistically significant gains confirmed via the Friedman test followed by the post-hoc Nemenyi test.
\par These results indicate that co-change entropy complements change entropy, significantly enhancing defect classification performance and underscoring its practical importance in defect prediction.
\end{abstract}



\maketitle
\section{Introduction and Motivation}
Software defect prediction has remained a central topic in software engineering for decades, as it enables the identification of potentially faulty modules, thereby guiding the efficient allocation of testing resources to enhance software quality. While a wide range of product and process metrics have been explored to build predictive models, the pursuit of more effective and reliable metrics remains ongoing. Notably, prior studies suggest that process metrics not only outperform product metrics in predictive performance \cite{bird2009does}\cite{bird2011don}\cite{rahman2013and}, but also offer practical advantages—they are easier to collect and are largely language-independent, thus improving their generalizability across diverse software projects \cite{majumder2022revisiting}.
\par The scattering of changes, also referred to as change entropy \cite{hassan2009predicting}, has been widely used as a process metric in defect prediction studies. Change entropy describes how dispersed changes are across the modules or files in a system during a specific time interval. Systems where changes are scattered across many files during a given period have been found to be more prone to defects in the future. Apart from the amount and patterns of changes made to a system and their dispersion, the patterns of co-changes—where two or more software entities are modified together in the same commit—have also been found to be associated with software quality \cite{silva2019co}\cite{silva2014assessing}\cite{d2009relationship}. However, despite the significant impact of co-change patterns on software quality and modularity, the effect of co-change scattering on defects has not been studied in depth, to the best of our knowledge. 
\par In this paper, similar to previous studies \cite{kumar2024prevalence}\cite{silva2014assessing}, we model the co-changes occurring in a software system as a graph and propose co-change graph entropy as a process metric to study its impact on defects in the system. Although co-change graph entropy may seem similar to change entropy, it is fundamentally different. To demonstrate how co-change graph entropy is computed and how it differs from change entropy, we consider a toy example where 12 commits ($C_1$, $C_2$, …, $C_{12}$) have been made in a system with four source code files (A, B, C, and D) during some time interval $T$. In the first 9 commits ($C_1$, $C_2$, …, $C_9$), files A and C are changed, while in commit $C_{10}$, files A and D are changed; in commit $C_{11}$, files C and D are changed; and in commit $C_{12}$, files B and D are changed.
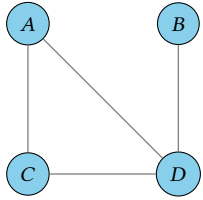
\begin{figure}[t]
  
    \centering
    \begin{adjustbox}{valign=t}
      \begin{tikzpicture}[
      node/.style={
        circle,
        draw,
        fill={skyblue},
        minimum size=3mm, 
        font=\small,
      },
      solidedge/.style={
        draw=grey,
        line width=0.5pt,
      },
      ]

      \node[node] (A) at (-1,1) {$A$}; 
      \node[node] (B) at (1,1) {$B$};
      \node[node] (C) at (-1,-1) {$C$};
      \node[node] (D) at (1,-1) {$D$};
     
      \draw[solidedge] (A) -- (D);
      \draw[solidedge] (B) -- (D);
      \draw[solidedge] (C) -- (D);
      \draw[solidedge] (A) -- (C);
     
      \end{tikzpicture}
    \end{adjustbox}
    \caption{Co-change graph based on the toy example}
    \Description{Co-change graph based on the toy example.}
    \label{fig:Co-change graph based on the toy example}
  \end{figure}
A corresponding co-change graph capturing the co-change relationships among the files in the above example is shown in Figure \ref{fig:Co-change graph based on the toy example}. 
\par Change entropy, as defined by Hassan \cite{hassan2009predicting}, first computes the probability of a file being changed in the time interval $T$, which is computed as the number of times a file $f$ has been changed during $T$
divided by the total number of file changes occurring in the system during the time period 
$T$. In the above example, the probability of each file being changed in the time period  $T$ is as follows: $p_A=10/24=0.416, p_B=1/24=0.0416, p_C=10/24=0.416, p_D=3/24=0.125$. The change entropy of the system is computed using Shannon's entropy formula as follows:.
\[
H(S) = - \sum_{k} p_k \log p_k
\] Where $p_k$ is the change probability of a file. Simple arithmetic can be used to compute the change entropy of the system in the toy example, which is 1.68551. Since defect proneness is to be investigated at the file level, the system's entropy in Hassan's model is attributed to the file level. As observed by Hassan, one of the best ways to assign the entropy measure at the file level is to multiply the system's entropy by the change probability of the file. In this way, each file is assigned a change complexity value as follows: $H_A=p_A*H(s)=0.673, H_B=p_B*H(s)=0.0673, H_C=p_C*H(s)=0.673, H_D=p_D*H(s)=0.202$. 
\par In contrast to change entropy, co-change graph entropy is based on the co-change probability of a node. The co-change probability of a node in a co-change graph is computed as the degree of the node divided by twice the number of edges in the graph. In this way, the co-change probability of the files, based on the co-change graph shown in Figure \ref{fig:Co-change graph based on the toy example}, can be calculated as follows:$p'_A=2/8=0.25, p'_B=1/8=0.125, p'_C=2/8=0.25, p'_D=3/8=0.375$. Corresponding co-change graph entropy $H'(S)$ for the system during time period $T$ can be computed as 1.90584. The co-change entropy assigned to each of the files can easily be computed as follows\footnote{Throughout this paper, "co-change entropy" and "change entropy" refer to the entropy assigned to a file, as our study focuses on defect prediction at the file level. Additionally, "module" and "source code file" are used interchangeably, as defect proneness is assessed at this level.}: $H'_A=p'_A*H'(s)=0.476, H'_B=p'_B*H'(s)=0.238, H'_C=p'_C*H'(s)=0.476, H'_D=p'_D*H'(s)=0.715$.
\par It is evident that files A and C have changed 10 times, resulting in significantly higher change entropy than files B and D. However, file D has a higher co-change entropy than A and B, indicating its greater co-change proneness. This highlights the fundamental difference between the two measures. Motivated by past studies highlighting the effectiveness of co-change patterns in assessing software quality and the importance of change entropy in defect prediction, we pose our first  research questions as follows:
\\\\
\fbox{\begin{minipage}{\dimexpr\columnwidth-2\fboxsep-2\fboxrule}
\textbf{RQ1:} \textit{To what extent does co-change entropy correlate with the defect count of software modules?}
\end{minipage}}
\\\\
In a typical defect prediction scenario, machine learning classifiers are trained on past data (e.g., process metrics of the previous release) and are used to predict defects in the future release of the software. Various process metrics have been proposed and found effective in predicting the defect proneness of modules and the number of defects in them \cite{rahman2013and}\cite{majumder2022revisiting}. Change entropy has been found to effectively complement other process metrics and improve the performance of defect prediction.
However, to investigate how our proposed co-change graph entropy complements other process metrics and how it performs in comparison to change entropy, we ask our second research question as follows:
\\\\
\fbox{\begin{minipage}{\dimexpr\columnwidth-2\fboxsep-2\fboxrule}
\textbf{RQ2:} \textit{Does co-change entropy complement standard process metrics and change entropy in improving defect prediction accuracy?}
\end{minipage}}
\\\\
To address our research questions, we conducted experiments on eight popular Apache projects from the SmartSHARK Dataset\cite{trautsch2021msr}. Our findings indicate that co-change entropy exhibits a significant correlation with defect count in source code files. Furthermore, co-change entropy proved highly effective in predicting defect proneness across various classifiers. In the majority of dataset and classifier combinations, co-change entropy effectively complemented standard process and change metrics, resulting in improved overall defect prediction performance.
\section{Related Work}
Software defect prediction relies heavily on metrics that capture different aspects of the development process. The two primary categories are product metrics, which assess code quality through measures like cyclomatic complexity, and process metrics, which leverage historical change data, such as lines of code added or deleted. Past research demonstrates the superior predictive power of process metrics over product metrics. Rahman and Devanbu \cite{rahman2013and} provided influential insights into this trend, which were later validated by large-scale studies such as those by Majumder et al. \cite{majumder2022revisiting}. Furthermore, Majumder et al. \cite{majumder2024less} showed that process metrics remain effective even in semi-supervised learning scenarios with limited labeled data.
Within process metrics, change-based measures—particularly those capturing patterns of past software version changes—have proven highly effective. Hassan \cite{hassan2009predicting} pioneered the use of change entropy at the file level to predict defects. Nagappan et al. \cite{nagappan2010change} identified "change bursts", or frequent changes within short intervals, as strong indicators of defect-prone modules. More recently, Wen et al. \cite{wen2018well} emphasized the importance of change sequences, employing Recurrent Neural Networks (RNNs) to model these sequences and significantly improve prediction accuracy.
\par The concept of co-change, or simultaneous modifications to multiple files, has been recognized as a valuable predictor of software defects. D’Ambros et al. \cite{d2009relationship} established foundational work by quantifying the impact of co-change frequency and defining metrics that demonstrated its utility in defect prediction. While D’Ambros et al. utilized numerous measures based on the raw count of co-changes per file, our approach diverges by first assessing the dispersion of co-changes at the system level. Subsequently, we derive a file-level metric that quantifies each file's contribution to the overall system’s co-change entropy. Notably, co-change dispersion and its attribution at the file level have not been used in prior studies, to the best of our knowledge. Kouroshfar \cite{kouroshfar2013studying} further explored co-change dispersion, identifying inter-subsystem co-changes as particularly defect-prone. However, Kouroshfar's analysis relied on simple count-based measures of co-changes within and across subsystems, lacking the nuanced, information-theoretic perspective offered by our graph entropy approach.
In contrast to prior work, our study introduces a novel approach using co-change graph entropy. This measure, based on information-theoretic entropy, quantifies co-change dispersion at the system level, moving beyond simple counts. We not only demonstrate the correlation between our proposed co-change entropy and software defects but also explore its ability to enhance the predictive power of other established process metrics. This comprehensive analysis highlights the unique contribution of our co-change entropy metric to improving defect prediction accuracy.
\section{Experimental setup}
This section outlines the datasets and experimental methodology utilized to answer our research questions.
\subsection{Dataset Description}
We conducted our experiments using the SmartSHARK dataset \cite{trautsch2021msr}, which is widely recognized and has been extensively used in prior research \cite{herbold2019costs} \cite{de2022studying}. This dataset undergoes regular updates and is integrated into a broader, well-structured data collection infrastructure\footnote{\url{https://smartshark.github.io/dbreleases/}}. Its usability and effectiveness have been demonstrated in numerous influential defect prediction studies \cite{herbold2022problems}\cite{trautsch2020static}.
\par We selected only the projects that have more than 2,000 defects and over 5,000 commits, resulting in a total of nine projects. As the release information for project Phoenix was ambiguous, pre-release changes could not be determined. Consequently, this project was excluded from the study. After this exclusion, our dataset consists of eight projects and 25 releases for defect analysis. The details of the dataset are provided in Table \ref{tab:projectstats}.
Our dataset complies with the guidelines laid down by Yatish et al. \cite{yatish2019mining}, ensuring that it includes a large number of fixed or closed issues, with a significant fraction of issue reports traced to the commits that resolve them. Furthermore, the dataset comprises projects of diverse nature and size, and the defect ratio across the studied releases also exhibits diversity.

\begin{table}[h!]
    \centering
    \scriptsize
    \caption{Dataset Description}
    \label{tab:projectstats}
    \begin{adjustbox}{max width=0.45\textwidth}
    \begin{tabular}{|c|c|c|c|c|c|c|}
\hline
\textit{\textbf{Project}}                   & \textit{\textbf{Release Type}} & \textit{\textbf{Release}} & \textit{\textbf{Edges}} & \textit{\textbf{Nodes}} & \textit{\textbf{Defects}} & \textit{\textbf{Defect Ratio}} \\ \hline
\multirow{2}{*}{\textit{\textbf{Derby}}}    & Training                       & 10.3.1.4                  & 159226                  & 1550                    & 309                       & 0.1653                         \\ \cline{2-7} 
                                            & Test                           & 10.5.1.1                  & 22589                   & 1485                    & 412                       & 0.240909                       \\ \hline
\multirow{5}{*}{\textit{\textbf{Activemq}}} & \multirow{4}{*}{Training}      & activemq-5.0.0            & 461396                  & 1291                    & 66                        & 0.046404                       \\ \cline{3-7} 
                                            &                                & activemq-5.1.0            & 4348                    & 294                     & 83                        & 0.248466                       \\ \cline{3-7} 
                                            &                                & activemq-5.2.0            & 4256                    & 297                     & 106                       & 0.302181                       \\ \cline{3-7} 
                                            &                                & activemq-5.3.0            & 13672                   & 539                     & 158                       & 0.227586                       \\ \cline{2-7} 
                                            & Test                           & activemq-5.5.0            & 1966322                 & 3259                    & 309                       & 0.082263                       \\ \hline
\multirow{4}{*}{\textit{\textbf{Pdfbox}}}   & \multirow{3}{*}{Training}      & 1.5.0                     & 1086                    & 203                     & 104                       & 0.435897                       \\ \cline{3-7} 
                                            &                                & 1.7.0                     & 9378                    & 416                     & 126                       & 0.273563                       \\ \cline{3-7} 
                                            &                                & 1.8.0                     & 159777                  & 1005                    & 76                        & 0.072277                       \\ \cline{2-7} 
                                            & Test                           & 2.0.0                     & 63155                   & 1559                    & 797                       & 0.489308                       \\ \hline
\multirow{4}{*}{\textit{\textbf{Pig}}}      & \multirow{3}{*}{Training}      & release-0.6.0             & 45588                   & 503                     & 99                        & 0.172962                       \\ \cline{3-7} 
                                            &                                & release-0.7.0             & 8853                    & 347                     & 51                        & 0.116343                       \\ \cline{3-7} 
                                            &                                & release-0.8.0             & 149036                  & 1034                    & 146                       & 0.117929                       \\ \cline{2-7} 
                                            & Test                           & release-0.9.0             & 13451                   & 474                     & 177                       & 0.282258                       \\ \hline
\multirow{2}{*}{\textit{\textbf{Kafka}}}    & Training                       & 0.10.0.0                  & 28379                   & 562                     & 202                       & 0.212544                       \\ \cline{2-7} 
                                            & Test                           & 0.11.0.0                  & 292651                  & 997                     & 186                       & 0.127617                       \\ \hline
\multirow{3}{*}{\textit{\textbf{Maven}}}    & \multirow{2}{*}{Training}      & maven-3.1.0               & 27025                   & 403                     & 178                       & 0.337408                       \\ \cline{3-7} 
                                            &                                & maven-3.3.9               & 63580                   & 538                     & 150                       & 0.188889                       \\ \cline{2-7} 
                                            & Test                           & maven-3.5.0               & 9970                    & 320                     & 114                       & 0.222222                       \\ \hline
\multirow{2}{*}{\textit{\textbf{Struts}}}   & Training                       & STRUTS\_2\_3\_28          & 3504646                 & 2724                    & 38                        & 0.012115                       \\ \cline{2-7} 
                                            & Test                           & STRUTS\_2\_3\_32          & 459284                  & 1034                    & 59                        & 0.04771                        \\ \hline
\multirow{3}{*}{\textit{\textbf{Nifi}}}     & \multirow{2}{*}{Training}      & nifi-0.5.0                & 106897                  & 592                     & 69                        & 0.08769                        \\ \cline{3-7} 
                                            &                                & nifi-0.6.0                & 5244                    & 197                     & 50                        & 0.200957                       \\ \cline{2-7} 
                                            & Test                           & nifi-0.7.0                & 97299                   & 1397                    & 97                        & 0.057123                       \\ \hline
\end{tabular}
    \end{adjustbox}
\end{table}
\vspace{-1em} 
\subsection{Set up to answer RQ1}
To assess whether co-change entropy can aid defect prediction, we analyze its correlation with post-release defects. We link defects to a release using the "affected version" field in the issue tracking system, as supported by past studies \cite{da2016framework}\cite{yatish2019mining}.
To collect the change data for a release  $R_i$, we follow this approach: if the release time for $R_i$ is $T_i$ and for the previous release it is $T_{i-1}$, then all commits made during $T_{i-1}$ and $T_i$ are assigned to release $R_i$. This follows past studies \cite{wen2018well}. We removed all commits that change more than 30 files (fatty commits) to ensure there is no noise in our data. Past research also suggests excluding such commits to avoid noise in the data \cite{xiao2014design}. 
\par For correlation analysis, we construct a co-change graph using change data for each release and compute co-change entropy at the file level. Specifically, for each source code file in a release, we obtain its co-change entropy and the number of post-release defects. We then aggregate this data across all studied releases of the project and compute the Spearman and Pearson correlation coefficients between these two variables.
\par One might question why we do not consider the number of times a pair of files has been co-changed and use a weighted co-change graph instead of an unweighted one. Upon closer examination, we find that as the frequency of co-changes among the same set of files increases, co-change entropy and change entropy tend to converge. This is evident in the toy example provided in the introduction. For a weighted graph, the co-change probability can be computed as $p_k=\frac{d(k)}{\sum_i d(i)}$, where $d(x)$ represents the weighted degree of node $x$ and ${\sum_i d(i)}$ is the weighted sum of the degrees of all nodes. However, a simple calculation shows that, in this case, the weighted co-change probability for each file effectively reduces to the file's change entropy, offering no additional benefit from a feature engineering perspective. Additionally, when edge weights are small, the co-change probability of the weighted graph approaches that of the unweighted graph. Recognizing these observations, we opted to use only the unweighted co-change graph in all our co-change entropy calculations. 
\subsection{Set up to answer RQ2} \label{set up for RQ2}
 The answer to RQ1 determines whether co-change entropy is useful for defect prediction. To evaluate its practical utility, we integrate it into machine learning classifiers alongside other metrics and analyze its impact on defect prediction accuracy. In addition to change entropy and co-change entropy-based file measures, we use process metrics used by Rahman and Devanbu \cite{rahman2013and}. To compare change entropy with co-change graph entropy, we define the following three sets of metrics for training machine learning classifiers on past releases and predicting defects in future releases.
\begin{itemize}[labelsep=0.5em,
                itemsep=1pt,    
                topsep=1pt] 
\item \textbf{P+C}: The metrics provided in Rahman $\&$ Devanbu's study, including change entropy (referred to as "changed code scattering" in their work).
\item \textbf{P+Co}: The metrics provided in Rahman $\&$ Devanbu's study, except that change entropy is replaced with co-change entropy.
\item \textbf{P+C+Co}: The metrics provided in Rahman $\&$ Devanbu's study, including change entropy + our proposed co-change entropy.
\end{itemize}
To investigate the effectiveness of co-change entropy in defect classification, we selected five widely used machine learning classifiers: Logistic Regression, Support Vector Machine, XGBoost, Random Forest, and Gradient Boosting. We used the SciKit-Learn \cite{pedregosa2011scikit} implementation of all these classifiers. The defect prediction dataset suffers from class imbalance. To address this, we apply SMOTE \cite{chawla2002smote}, a widely used class balancing technique in software engineering research \cite{majumder2022revisiting}\cite{wang2013using}.
\par In a typical cross-version defect classification task utilizing process metrics, software development process quality metrics for software modules are collected from previous releases and used to train a binary classifier that predicts defect proneness in a future release. Generally, for releases $R_1, R_2, R_3, R_4, \dots, R_{n-1}$, metric values are collected, and the defect proneness of modules is predicted for release $R_n$.
\par For each classifier and each project, we train three models using the P+C, P+Co, and P+C+Co sets of metrics. The process metrics from training releases are used to evaluate predictive performance on the test release. To assess classifier performance, we use AUROC, F1-score, MCC (Matthews Correlation Coefficient), Precision, and Recall. These metrics are widely used for binary classification and have been extensively used in defect classification research \cite{gong2021revisiting}\cite{rahman2013and}\cite{tantithamthavorn2018impact}.

\begin{table}[h!]
    \centering
    \scriptsize
    \caption{Correlation analysis with Bug Count}
    \label{tab:CORRELATION TABLE}
    \begin{adjustbox}{max width=0.45\textwidth}
    \begin{tabular}{|c|cccc|cccc|}
\hline
\textit{\textbf{}}         & \multicolumn{4}{c|}{\textit{\textbf{Change Entropy}}}                                                                                                                        & \multicolumn{4}{c|}{\textit{\textbf{Co\_Change\_Entropy}}}                                                                                                                   \\ \hline
\textit{\textbf{Project}}  & \multicolumn{1}{c|}{\textit{\textbf{P\_Corr}}} & \multicolumn{1}{c|}{\textit{\textbf{P\_pval}}} & \multicolumn{1}{c|}{\textit{\textbf{S\_Corr}}} & \textit{\textbf{S\_pval}} & \multicolumn{1}{c|}{\textit{\textbf{P\_Corr}}} & \multicolumn{1}{c|}{\textit{\textbf{P\_pval}}} & \multicolumn{1}{c|}{\textit{\textbf{S\_Corr}}} & \textit{\textbf{S\_pval}} \\ \hline
\textit{\textbf{Derby}}    & \multicolumn{1}{c|}{0.779}                     & \multicolumn{1}{c|}{0.00E+00}                  & \multicolumn{1}{c|}{0.567}                     & 1.78E-204                 & \multicolumn{1}{c|}{0.461}                     & \multicolumn{1}{c|}{1.44E-126}                 & \multicolumn{1}{c|}{0.25}                      & 2.05E-35                  \\ \hline
\textit{\textbf{Activemq}} & \multicolumn{1}{c|}{0.548}                     & \multicolumn{1}{c|}{3.60E-129}                 & \multicolumn{1}{c|}{0.311}                     & 4.23E-38                  & \multicolumn{1}{c|}{0.516}                     & \multicolumn{1}{c|}{3.94E-112}                 & \multicolumn{1}{c|}{0.283}                     & 1.37E-31                  \\ \hline
\textit{\textbf{Pdfbox}}   & \multicolumn{1}{c|}{0.385}                     & \multicolumn{1}{c|}{1.13E-71}                  & \multicolumn{1}{c|}{0.411}                     & 1.55E-82                  & \multicolumn{1}{c|}{0.125}                     & \multicolumn{1}{c|}{2.01E-08}                  & \multicolumn{1}{c|}{0.196}                     & 7.31E-19                  \\ \hline
\textit{\textbf{Pig}}      & \multicolumn{1}{c|}{0.683}                     & \multicolumn{1}{c|}{1.82E-151}                 & \multicolumn{1}{c|}{0.445}                     & 2.32E-54                  & \multicolumn{1}{c|}{0.358}                     & \multicolumn{1}{c|}{1.93E-34}                  & \multicolumn{1}{c|}{0.145}                     & 1.41E-06                  \\ \hline
\textit{\textbf{Kafka}}    & \multicolumn{1}{c|}{0.729}                     & \multicolumn{1}{c|}{4.58E-166}                 & \multicolumn{1}{c|}{0.522}                     & 9.05E-71                  & \multicolumn{1}{c|}{0.314}                     & \multicolumn{1}{c|}{2.93E-24}                  & \multicolumn{1}{c|}{0.13}                      & 3.97E-05                  \\ \hline
\textit{\textbf{Maven}}    & \multicolumn{1}{c|}{0.637}                     & \multicolumn{1}{c|}{1.50E-86}                  & \multicolumn{1}{c|}{0.405}                     & 5.65E-31                  & \multicolumn{1}{c|}{0.343}                     & \multicolumn{1}{c|}{3.63E-22}                  & \multicolumn{1}{c|}{0.166}                     & 4.89E-06                  \\ \hline
\textit{\textbf{Struts}}   & \multicolumn{1}{c|}{0.467}                     & \multicolumn{1}{c|}{1.27E-30}                  & \multicolumn{1}{c|}{0.321}                     & 2.05E-14                  & \multicolumn{1}{c|}{0.542}                     & \multicolumn{1}{c|}{1.47E-42}                  & \multicolumn{1}{c|}{0.346}                     & 1.16E-16                  \\ \hline
\textit{\textbf{Nifi}}     & \multicolumn{1}{c|}{0.337}                     & \multicolumn{1}{c|}{3.96E-28}                  & \multicolumn{1}{c|}{0.359}                     & 5.78E-32                  & \multicolumn{1}{c|}{0.143}                     & \multicolumn{1}{c|}{5.32E-06}                  & \multicolumn{1}{c|}{0.006}                     & 8.54E-01                  \\ \hline
\end{tabular}
    \end{adjustbox}
\begin{minipage}{0.45\textwidth}
\centering
        \scriptsize \textbf{Note:} P -> Pearson, S -> Spearman, Corr -> Correlation, pval -> p-value.
    \end{minipage}
\end{table}
\begin{table*}[h!]
    \centering
    \scriptsize
    \caption{Defect Classification Results}
    \label{tab:prediction}
    \begin{adjustbox}{max width=\textwidth}
\begin{tabular}{|cc|ccc|ccc|ccc|ccc|ccc|}
\hline
\multicolumn{2}{|c|}{\textit{\textbf{Classifier}}}                                              & \multicolumn{3}{c|}{\textit{\textbf{Logistic Regression}}}                                                                  & \multicolumn{3}{c|}{\textit{\textbf{Support Vector Machine}}}                                                               & \multicolumn{3}{c|}{\textit{\textbf{Random Forest}}}                                                                        & \multicolumn{3}{c|}{\textit{\textbf{XGBoost}}}                                                                              & \multicolumn{3}{c|}{\textit{\textbf{Gradient Boosting}}}                                                                    \\ \hline
\multicolumn{1}{|c|}{\textit{\textbf{Project}}}                   & \textit{\textbf{Metrics}}   & \multicolumn{1}{c|}{\textit{\textbf{P + C}}} & \multicolumn{1}{c|}{\textit{\textbf{P + Co}}} & \textit{\textbf{P + C + Co}} & \multicolumn{1}{c|}{\textit{\textbf{P + C}}} & \multicolumn{1}{c|}{\textit{\textbf{P + Co}}} & \textit{\textbf{P + C + Co}} & \multicolumn{1}{c|}{\textit{\textbf{P + C}}} & \multicolumn{1}{c|}{\textit{\textbf{P + Co}}} & \textit{\textbf{P + C + Co}} & \multicolumn{1}{c|}{\textit{\textbf{P + C}}} & \multicolumn{1}{c|}{\textit{\textbf{P + Co}}} & \textit{\textbf{P + C + Co}} & \multicolumn{1}{c|}{\textit{\textbf{P + C}}} & \multicolumn{1}{c|}{\textit{\textbf{P + Co}}} & \textit{\textbf{P + C + Co}} \\ \hline
\multicolumn{1}{|c|}{\multirow{5}{*}{\textit{\textbf{Derby}}}}    & \textit{\textbf{AUROC}}     & \multicolumn{1}{c|}{0.77}                    & \multicolumn{1}{c|}{0.763}                    & 0.77                         & \multicolumn{1}{c|}{0.731}                   & \multicolumn{1}{c|}{0.718}                    & 0.724                        & \multicolumn{1}{c|}{0.789}                   & \multicolumn{1}{c|}{0.799}                    & 0.792                        & \multicolumn{1}{c|}{0.785}                   & \multicolumn{1}{c|}{0.773}                    & 0.773                        & \multicolumn{1}{c|}{0.787}                   & \multicolumn{1}{c|}{0.797}                    & 0.8                          \\ \cline{2-17} 
\multicolumn{1}{|c|}{}                                            & \textit{\textbf{F1-Score}}  & \multicolumn{1}{c|}{0.649}                   & \multicolumn{1}{c|}{0.634}                    & 0.65                         & \multicolumn{1}{c|}{0.594}                   & \multicolumn{1}{c|}{0.593}                    & 0.586                        & \multicolumn{1}{c|}{0.685}                   & \multicolumn{1}{c|}{0.688}                    & 0.686                        & \multicolumn{1}{c|}{0.695}                   & \multicolumn{1}{c|}{0.673}                    & 0.673                        & \multicolumn{1}{c|}{0.688}                   & \multicolumn{1}{c|}{0.69}                     & 0.698                        \\ \cline{2-17} 
\multicolumn{1}{|c|}{}                                            & \textit{\textbf{MCC}}       & \multicolumn{1}{c|}{0.447}                   & \multicolumn{1}{c|}{0.419}                    & 0.451                        & \multicolumn{1}{c|}{0.389}                   & \multicolumn{1}{c|}{0.394}                    & 0.387                        & \multicolumn{1}{c|}{0.414}                   & \multicolumn{1}{c|}{0.449}                    & 0.431                        & \multicolumn{1}{c|}{0.409}                   & \multicolumn{1}{c|}{0.371}                    & 0.371                        & \multicolumn{1}{c|}{0.386}                   & \multicolumn{1}{c|}{0.431}                    & 0.417                        \\ \cline{2-17} 
\multicolumn{1}{|c|}{}                                            & \textit{\textbf{Precision}} & \multicolumn{1}{c|}{0.823}                   & \multicolumn{1}{c|}{0.802}                    & 0.827                        & \multicolumn{1}{c|}{0.804}                   & \multicolumn{1}{c|}{0.811}                    & 0.809                        & \multicolumn{1}{c|}{0.738}                   & \multicolumn{1}{c|}{0.779}                    & 0.758                        & \multicolumn{1}{c|}{0.72}                    & \multicolumn{1}{c|}{0.703}                    & 0.703                        & \multicolumn{1}{c|}{0.703}                   & \multicolumn{1}{c|}{0.754}                    & 0.726                        \\ \cline{2-17} 
\multicolumn{1}{|c|}{}                                            & \textit{\textbf{Recall}}    & \multicolumn{1}{c|}{0.535}                   & \multicolumn{1}{c|}{0.524}                    & 0.535                        & \multicolumn{1}{c|}{0.471}                   & \multicolumn{1}{c|}{0.468}                    & 0.46                         & \multicolumn{1}{c|}{0.639}                   & \multicolumn{1}{c|}{0.616}                    & 0.627                        & \multicolumn{1}{c|}{0.672}                   & \multicolumn{1}{c|}{0.646}                    & 0.646                        & \multicolumn{1}{c|}{0.674}                   & \multicolumn{1}{c|}{0.636}                    & 0.672                        \\ \hline
\multicolumn{1}{|c|}{\multirow{5}{*}{\textit{\textbf{Activemq}}}} & \textit{\textbf{AUROC}}     & \multicolumn{1}{c|}{0.771}                   & \multicolumn{1}{c|}{0.777}                    & 0.778                        & \multicolumn{1}{c|}{0.691}                   & \multicolumn{1}{c|}{0.708}                    & 0.709                        & \multicolumn{1}{c|}{0.763}                   & \multicolumn{1}{c|}{0.747}                    & 0.775                        & \multicolumn{1}{c|}{0.759}                   & \multicolumn{1}{c|}{0.772}                    & 0.759                        & \multicolumn{1}{c|}{0.717}                   & \multicolumn{1}{c|}{0.675}                    & 0.687                        \\ \cline{2-17} 
\multicolumn{1}{|c|}{}                                            & \textit{\textbf{F1-Score}}  & \multicolumn{1}{c|}{0.792}                   & \multicolumn{1}{c|}{0.796}                    & 0.801                        & \multicolumn{1}{c|}{0.785}                   & \multicolumn{1}{c|}{0.793}                    & 0.792                        & \multicolumn{1}{c|}{0.787}                   & \multicolumn{1}{c|}{0.802}                    & 0.791                        & \multicolumn{1}{c|}{0.804}                   & \multicolumn{1}{c|}{0.812}                    & 0.792                        & \multicolumn{1}{c|}{0.758}                   & \multicolumn{1}{c|}{0.752}                    & 0.75                         \\ \cline{2-17} 
\multicolumn{1}{|c|}{}                                            & \textit{\textbf{MCC}}       & \multicolumn{1}{c|}{0.323}                   & \multicolumn{1}{c|}{0.339}                    & 0.355                        & \multicolumn{1}{c|}{0.272}                   & \multicolumn{1}{c|}{0.291}                    & 0.292                        & \multicolumn{1}{c|}{0.358}                   & \multicolumn{1}{c|}{0.353}                    & 0.358                        & \multicolumn{1}{c|}{0.381}                   & \multicolumn{1}{c|}{0.391}                    & 0.331                        & \multicolumn{1}{c|}{0.263}                   & \multicolumn{1}{c|}{0.2}                      & 0.213                        \\ \cline{2-17} 
\multicolumn{1}{|c|}{}                                            & \textit{\textbf{Precision}} & \multicolumn{1}{c|}{0.733}                   & \multicolumn{1}{c|}{0.738}                    & 0.743                        & \multicolumn{1}{c|}{0.712}                   & \multicolumn{1}{c|}{0.714}                    & 0.715                        & \multicolumn{1}{c|}{0.76}                    & \multicolumn{1}{c|}{0.739}                    & 0.756                        & \multicolumn{1}{c|}{0.756}                   & \multicolumn{1}{c|}{0.751}                    & 0.738                        & \multicolumn{1}{c|}{0.729}                   & \multicolumn{1}{c|}{0.702}                    & 0.709                        \\ \cline{2-17} 
\multicolumn{1}{|c|}{}                                            & \textit{\textbf{Recall}}    & \multicolumn{1}{c|}{0.861}                   & \multicolumn{1}{c|}{0.865}                    & 0.868                        & \multicolumn{1}{c|}{0.875}                   & \multicolumn{1}{c|}{0.891}                    & 0.888                        & \multicolumn{1}{c|}{0.815}                   & \multicolumn{1}{c|}{0.878}                    & 0.828                        & \multicolumn{1}{c|}{0.858}                   & \multicolumn{1}{c|}{0.884}                    & 0.855                        & \multicolumn{1}{c|}{0.789}                   & \multicolumn{1}{c|}{0.809}                    & 0.795                        \\ \hline
\multicolumn{1}{|c|}{\multirow{5}{*}{\textit{\textbf{Pdfbox}}}}   & \textit{\textbf{AUROC}}     & \multicolumn{1}{c|}{0.719}                   & \multicolumn{1}{c|}{0.725}                    & 0.72                         & \multicolumn{1}{c|}{0.54}                    & \multicolumn{1}{c|}{0.53}                     & 0.539                        & \multicolumn{1}{c|}{0.774}                   & \multicolumn{1}{c|}{0.753}                    & 0.792                        & \multicolumn{1}{c|}{0.739}                   & \multicolumn{1}{c|}{0.751}                    & 0.749                        & \multicolumn{1}{c|}{0.725}                   & \multicolumn{1}{c|}{0.777}                    & 0.709                        \\ \cline{2-17} 
\multicolumn{1}{|c|}{}                                            & \textit{\textbf{F1-Score}}  & \multicolumn{1}{c|}{0.756}                   & \multicolumn{1}{c|}{0.762}                    & 0.761                        & \multicolumn{1}{c|}{0.738}                   & \multicolumn{1}{c|}{0.756}                    & 0.753                        & \multicolumn{1}{c|}{0.777}                   & \multicolumn{1}{c|}{0.776}                    & 0.795                        & \multicolumn{1}{c|}{0.769}                   & \multicolumn{1}{c|}{0.776}                    & 0.782                        & \multicolumn{1}{c|}{0.766}                   & \multicolumn{1}{c|}{0.784}                    & 0.757                        \\ \cline{2-17} 
\multicolumn{1}{|c|}{}                                            & \textit{\textbf{MCC}}       & \multicolumn{1}{c|}{0.217}                   & \multicolumn{1}{c|}{0.231}                    & 0.229                        & \multicolumn{1}{c|}{0.202}                   & \multicolumn{1}{c|}{0.224}                    & 0.224                        & \multicolumn{1}{c|}{0.28}                    & \multicolumn{1}{c|}{0.272}                    & 0.378                        & \multicolumn{1}{c|}{0.231}                   & \multicolumn{1}{c|}{0.266}                    & 0.295                        & \multicolumn{1}{c|}{0.249}                   & \multicolumn{1}{c|}{0.3}                      & 0.265                        \\ \cline{2-17} 
\multicolumn{1}{|c|}{}                                            & \textit{\textbf{Precision}} & \multicolumn{1}{c|}{0.658}                   & \multicolumn{1}{c|}{0.658}                    & 0.659                        & \multicolumn{1}{c|}{0.663}                   & \multicolumn{1}{c|}{0.66}                     & 0.663                        & \multicolumn{1}{c|}{0.664}                   & \multicolumn{1}{c|}{0.661}                    & 0.704                        & \multicolumn{1}{c|}{0.649}                   & \multicolumn{1}{c|}{0.656}                    & 0.664                        & \multicolumn{1}{c|}{0.662}                   & \multicolumn{1}{c|}{0.66}                     & 0.681                        \\ \cline{2-17} 
\multicolumn{1}{|c|}{}                                            & \textit{\textbf{Recall}}    & \multicolumn{1}{c|}{0.888}                   & \multicolumn{1}{c|}{0.907}                    & 0.9                          & \multicolumn{1}{c|}{0.833}                   & \multicolumn{1}{c|}{0.884}                    & 0.871                        & \multicolumn{1}{c|}{0.934}                   & \multicolumn{1}{c|}{0.938}                    & 0.912                        & \multicolumn{1}{c|}{0.943}                   & \multicolumn{1}{c|}{0.95}                     & 0.95                         & \multicolumn{1}{c|}{0.909}                   & \multicolumn{1}{c|}{0.965}                    & 0.852                        \\ \hline
\multicolumn{1}{|c|}{\multirow{5}{*}{\textit{\textbf{Pig}}}}      & \textit{\textbf{AUROC}}     & \multicolumn{1}{c|}{0.621}                   & \multicolumn{1}{c|}{0.601}                    & 0.621                        & \multicolumn{1}{c|}{0.595}                   & \multicolumn{1}{c|}{0.586}                    & 0.602                        & \multicolumn{1}{c|}{0.635}                   & \multicolumn{1}{c|}{0.607}                    & 0.617                        & \multicolumn{1}{c|}{0.613}                   & \multicolumn{1}{c|}{0.609}                    & 0.636                        & \multicolumn{1}{c|}{0.658}                   & \multicolumn{1}{c|}{0.624}                    & 0.649                        \\ \cline{2-17} 
\multicolumn{1}{|c|}{}                                            & \textit{\textbf{F1-Score}}  & \multicolumn{1}{c|}{0.68}                    & \multicolumn{1}{c|}{0.682}                    & 0.68                         & \multicolumn{1}{c|}{0.686}                   & \multicolumn{1}{c|}{0.69}                     & 0.688                        & \multicolumn{1}{c|}{0.696}                   & \multicolumn{1}{c|}{0.687}                    & 0.689                        & \multicolumn{1}{c|}{0.725}                   & \multicolumn{1}{c|}{0.708}                    & 0.712                        & \multicolumn{1}{c|}{0.681}                   & \multicolumn{1}{c|}{0.691}                    & 0.694                        \\ \cline{2-17} 
\multicolumn{1}{|c|}{}                                            & \textit{\textbf{MCC}}       & \multicolumn{1}{c|}{0.105}                   & \multicolumn{1}{c|}{0.115}                    & 0.105                        & \multicolumn{1}{c|}{0.107}                   & \multicolumn{1}{c|}{0.127}                    & 0.131                        & \multicolumn{1}{c|}{0.067}                   & \multicolumn{1}{c|}{0.073}                    & 0.054                        & \multicolumn{1}{c|}{0.148}                   & \multicolumn{1}{c|}{0.144}                    & 0.118                        & \multicolumn{1}{c|}{0.026}                   & \multicolumn{1}{c|}{0.08}                     & 0.071                        \\ \cline{2-17} 
\multicolumn{1}{|c|}{}                                            & \textit{\textbf{Precision}} & \multicolumn{1}{c|}{0.682}                   & \multicolumn{1}{c|}{0.686}                    & 0.682                        & \multicolumn{1}{c|}{0.682}                   & \multicolumn{1}{c|}{0.69}                     & 0.692                        & \multicolumn{1}{c|}{0.665}                   & \multicolumn{1}{c|}{0.668}                    & 0.661                        & \multicolumn{1}{c|}{0.689}                   & \multicolumn{1}{c|}{0.692}                    & 0.681                        & \multicolumn{1}{c|}{0.653}                   & \multicolumn{1}{c|}{0.67}                     & 0.667                        \\ \cline{2-17} 
\multicolumn{1}{|c|}{}                                            & \textit{\textbf{Recall}}    & \multicolumn{1}{c|}{0.678}                   & \multicolumn{1}{c|}{0.678}                    & 0.678                        & \multicolumn{1}{c|}{0.69}                    & \multicolumn{1}{c|}{0.69}                     & 0.684                        & \multicolumn{1}{c|}{0.73}                    & \multicolumn{1}{c|}{0.707}                    & 0.718                        & \multicolumn{1}{c|}{0.764}                   & \multicolumn{1}{c|}{0.724}                    & 0.747                        & \multicolumn{1}{c|}{0.713}                   & \multicolumn{1}{c|}{0.713}                    & 0.724                        \\ \hline
\multicolumn{1}{|c|}{\multirow{5}{*}{\textit{\textbf{Kafka}}}}    & \textit{\textbf{AUROC}}     & \multicolumn{1}{c|}{0.674}                   & \multicolumn{1}{c|}{0.678}                    & 0.679                        & \multicolumn{1}{c|}{0.632}                   & \multicolumn{1}{c|}{0.632}                    & 0.634                        & \multicolumn{1}{c|}{0.677}                   & \multicolumn{1}{c|}{0.71}                     & 0.734                        & \multicolumn{1}{c|}{0.625}                   & \multicolumn{1}{c|}{0.682}                    & 0.682                        & \multicolumn{1}{c|}{0.646}                   & \multicolumn{1}{c|}{0.683}                    & 0.681                        \\ \cline{2-17} 
\multicolumn{1}{|c|}{}                                            & \textit{\textbf{F1-Score}}  & \multicolumn{1}{c|}{0.544}                   & \multicolumn{1}{c|}{0.536}                    & 0.55                         & \multicolumn{1}{c|}{0.577}                   & \multicolumn{1}{c|}{0.587}                    & 0.573                        & \multicolumn{1}{c|}{0.613}                   & \multicolumn{1}{c|}{0.632}                    & 0.634                        & \multicolumn{1}{c|}{0.619}                   & \multicolumn{1}{c|}{0.663}                    & 0.663                        & \multicolumn{1}{c|}{0.623}                   & \multicolumn{1}{c|}{0.662}                    & 0.65                         \\ \cline{2-17} 
\multicolumn{1}{|c|}{}                                            & \textit{\textbf{MCC}}       & \multicolumn{1}{c|}{0.211}                   & \multicolumn{1}{c|}{0.199}                    & 0.224                        & \multicolumn{1}{c|}{0.162}                   & \multicolumn{1}{c|}{0.178}                    & 0.15                         & \multicolumn{1}{c|}{0.227}                   & \multicolumn{1}{c|}{0.283}                    & 0.275                        & \multicolumn{1}{c|}{0.145}                   & \multicolumn{1}{c|}{0.257}                    & 0.257                        & \multicolumn{1}{c|}{0.19}                    & \multicolumn{1}{c|}{0.273}                    & 0.259                        \\ \cline{2-17} 
\multicolumn{1}{|c|}{}                                            & \textit{\textbf{Precision}} & \multicolumn{1}{c|}{0.632}                   & \multicolumn{1}{c|}{0.625}                    & 0.641                        & \multicolumn{1}{c|}{0.578}                   & \multicolumn{1}{c|}{0.585}                    & 0.571                        & \multicolumn{1}{c|}{0.609}                   & \multicolumn{1}{c|}{0.644}                    & 0.635                        & \multicolumn{1}{c|}{0.551}                   & \multicolumn{1}{c|}{0.596}                    & 0.596                        & \multicolumn{1}{c|}{0.576}                   & \multicolumn{1}{c|}{0.61}                     & 0.609                        \\ \cline{2-17} 
\multicolumn{1}{|c|}{}                                            & \textit{\textbf{Recall}}    & \multicolumn{1}{c|}{0.478}                   & \multicolumn{1}{c|}{0.469}                    & 0.482                        & \multicolumn{1}{c|}{0.576}                   & \multicolumn{1}{c|}{0.588}                    & 0.576                        & \multicolumn{1}{c|}{0.616}                   & \multicolumn{1}{c|}{0.62}                     & 0.633                        & \multicolumn{1}{c|}{0.706}                   & \multicolumn{1}{c|}{0.747}                    & 0.747                        & \multicolumn{1}{c|}{0.678}                   & \multicolumn{1}{c|}{0.722}                    & 0.698                        \\ \hline
\multicolumn{1}{|c|}{\multirow{5}{*}{\textit{\textbf{Maven}}}}    & \textit{\textbf{AUROC}}     & \multicolumn{1}{c|}{0.66}                    & \multicolumn{1}{c|}{0.648}                    & 0.668                        & \multicolumn{1}{c|}{0.622}                   & \multicolumn{1}{c|}{0.597}                    & 0.611                        & \multicolumn{1}{c|}{0.646}                   & \multicolumn{1}{c|}{0.658}                    & 0.668                        & \multicolumn{1}{c|}{0.599}                   & \multicolumn{1}{c|}{0.616}                    & 0.638                        & \multicolumn{1}{c|}{0.636}                   & \multicolumn{1}{c|}{0.62}                     & 0.641                        \\ \cline{2-17} 
\multicolumn{1}{|c|}{}                                            & \textit{\textbf{F1-Score}}  & \multicolumn{1}{c|}{0.54}                    & \multicolumn{1}{c|}{0.475}                    & 0.5                          & \multicolumn{1}{c|}{0.442}                   & \multicolumn{1}{c|}{0.442}                    & 0.462                        & \multicolumn{1}{c|}{0.424}                   & \multicolumn{1}{c|}{0.293}                    & 0.4                          & \multicolumn{1}{c|}{0.4}                     & \multicolumn{1}{c|}{0.265}                    & 0.376                        & \multicolumn{1}{c|}{0.467}                   & \multicolumn{1}{c|}{0.349}                    & 0.378                        \\ \cline{2-17} 
\multicolumn{1}{|c|}{}                                            & \textit{\textbf{MCC}}       & \multicolumn{1}{c|}{0.298}                   & \multicolumn{1}{c|}{0.226}                    & 0.244                        & \multicolumn{1}{c|}{0.199}                   & \multicolumn{1}{c|}{0.199}                    & 0.211                        & \multicolumn{1}{c|}{0.243}                   & \multicolumn{1}{c|}{0.193}                    & 0.235                        & \multicolumn{1}{c|}{0.175}                   & \multicolumn{1}{c|}{0.023}                    & 0.163                        & \multicolumn{1}{c|}{0.207}                   & \multicolumn{1}{c|}{0.081}                    & 0.115                        \\ \cline{2-17} 
\multicolumn{1}{|c|}{}                                            & \textit{\textbf{Precision}} & \multicolumn{1}{c|}{0.531}                   & \multicolumn{1}{c|}{0.5}                      & 0.5                          & \multicolumn{1}{c|}{0.49}                    & \multicolumn{1}{c|}{0.49}                     & 0.491                        & \multicolumn{1}{c|}{0.568}                   & \multicolumn{1}{c|}{0.6}                      & 0.576                        & \multicolumn{1}{c|}{0.488}                   & \multicolumn{1}{c|}{0.361}                    & 0.487                        & \multicolumn{1}{c|}{0.483}                   & \multicolumn{1}{c|}{0.404}                    & 0.429                        \\ \cline{2-17} 
\multicolumn{1}{|c|}{}                                            & \textit{\textbf{Recall}}    & \multicolumn{1}{c|}{0.548}                   & \multicolumn{1}{c|}{0.452}                    & 0.5                          & \multicolumn{1}{c|}{0.403}                   & \multicolumn{1}{c|}{0.403}                    & 0.435                        & \multicolumn{1}{c|}{0.339}                   & \multicolumn{1}{c|}{0.194}                    & 0.306                        & \multicolumn{1}{c|}{0.339}                   & \multicolumn{1}{c|}{0.21}                     & 0.306                        & \multicolumn{1}{c|}{0.452}                   & \multicolumn{1}{c|}{0.306}                    & 0.339                        \\ \hline
\multicolumn{1}{|c|}{\multirow{5}{*}{\textit{\textbf{Struts}}}}   & \textit{\textbf{AUROC}}     & \multicolumn{1}{c|}{0.499}                   & \multicolumn{1}{c|}{0.573}                    & 0.58                         & \multicolumn{1}{c|}{0.727}                   & \multicolumn{1}{c|}{0.749}                    & 0.743                        & \multicolumn{1}{c|}{0.705}                   & \multicolumn{1}{c|}{0.722}                    & 0.744                        & \multicolumn{1}{c|}{0.649}                   & \multicolumn{1}{c|}{0.685}                    & 0.685                        & \multicolumn{1}{c|}{0.751}                   & \multicolumn{1}{c|}{0.702}                    & 0.7                          \\ \cline{2-17} 
\multicolumn{1}{|c|}{}                                            & \textit{\textbf{F1-Score}}  & \multicolumn{1}{c|}{0.703}                   & \multicolumn{1}{c|}{0.739}                    & 0.734                        & \multicolumn{1}{c|}{0.684}                   & \multicolumn{1}{c|}{0.688}                    & 0.688                        & \multicolumn{1}{c|}{0.613}                   & \multicolumn{1}{c|}{0.634}                    & 0.607                        & \multicolumn{1}{c|}{0.613}                   & \multicolumn{1}{c|}{0.741}                    & 0.741                        & \multicolumn{1}{c|}{0.711}                   & \multicolumn{1}{c|}{0.724}                    & 0.708                        \\ \cline{2-17} 
\multicolumn{1}{|c|}{}                                            & \textit{\textbf{MCC}}       & \multicolumn{1}{c|}{-0.25}                   & \multicolumn{1}{c|}{-0.228}                   & -0.236                       & \multicolumn{1}{c|}{0.426}                   & \multicolumn{1}{c|}{0.448}                    & 0.448                        & \multicolumn{1}{c|}{0.361}                   & \multicolumn{1}{c|}{0.27}                     & 0.418                        & \multicolumn{1}{c|}{0.197}                   & \multicolumn{1}{c|}{0.12}                     & 0.12                         & \multicolumn{1}{c|}{0.265}                   & \multicolumn{1}{c|}{0.151}                    & 0.141                        \\ \cline{2-17} 
\multicolumn{1}{|c|}{}                                            & \textit{\textbf{Precision}} & \multicolumn{1}{c|}{0.639}                   & \multicolumn{1}{c|}{0.654}                    & 0.651                        & \multicolumn{1}{c|}{0.931}                   & \multicolumn{1}{c|}{0.947}                    & 0.947                        & \multicolumn{1}{c|}{0.92}                    & \multicolumn{1}{c|}{0.836}                    & 0.978                        & \multicolumn{1}{c|}{0.794}                   & \multicolumn{1}{c|}{0.724}                    & 0.724                        & \multicolumn{1}{c|}{0.8}                     & \multicolumn{1}{c|}{0.74}                     & 0.739                        \\ \cline{2-17} 
\multicolumn{1}{|c|}{}                                            & \textit{\textbf{Recall}}    & \multicolumn{1}{c|}{0.78}                    & \multicolumn{1}{c|}{0.85}                     & 0.84                         & \multicolumn{1}{c|}{0.54}                    & \multicolumn{1}{c|}{0.54}                     & 0.54                         & \multicolumn{1}{c|}{0.46}                    & \multicolumn{1}{c|}{0.51}                     & 0.44                         & \multicolumn{1}{c|}{0.5}                     & \multicolumn{1}{c|}{0.76}                     & 0.76                         & \multicolumn{1}{c|}{0.64}                    & \multicolumn{1}{c|}{0.71}                     & 0.68                         \\ \hline
\multicolumn{1}{|c|}{\multirow{5}{*}{\textit{\textbf{Nifi}}}}     & \textit{\textbf{AUROC}}     & \multicolumn{1}{c|}{0.67}                    & \multicolumn{1}{c|}{0.7}                      & 0.703                        & \multicolumn{1}{c|}{0.586}                   & \multicolumn{1}{c|}{0.553}                    & 0.582                        & \multicolumn{1}{c|}{0.687}                   & \multicolumn{1}{c|}{0.666}                    & 0.711                        & \multicolumn{1}{c|}{0.625}                   & \multicolumn{1}{c|}{0.685}                    & 0.69                         & \multicolumn{1}{c|}{0.555}                   & \multicolumn{1}{c|}{0.616}                    & 0.641                        \\ \cline{2-17} 
\multicolumn{1}{|c|}{}                                            & \textit{\textbf{F1-Score}}  & \multicolumn{1}{c|}{0.449}                   & \multicolumn{1}{c|}{0.5}                      & 0.498                        & \multicolumn{1}{c|}{0.479}                   & \multicolumn{1}{c|}{0.471}                    & 0.483                        & \multicolumn{1}{c|}{0.544}                   & \multicolumn{1}{c|}{0.515}                    & 0.55                         & \multicolumn{1}{c|}{0.469}                   & \multicolumn{1}{c|}{0.49}                     & 0.498                        & \multicolumn{1}{c|}{0.466}                   & \multicolumn{1}{c|}{0.494}                    & 0.473                        \\ \cline{2-17} 
\multicolumn{1}{|c|}{}                                            & \textit{\textbf{MCC}}       & \multicolumn{1}{c|}{0.17}                    & \multicolumn{1}{c|}{0.232}                    & 0.227                        & \multicolumn{1}{c|}{0.194}                   & \multicolumn{1}{c|}{0.178}                    & 0.195                        & \multicolumn{1}{c|}{0.273}                   & \multicolumn{1}{c|}{0.21}                     & 0.288                        & \multicolumn{1}{c|}{0.09}                    & \multicolumn{1}{c|}{0.138}                    & 0.163                        & \multicolumn{1}{c|}{0.101}                   & \multicolumn{1}{c|}{0.172}                    & 0.116                        \\ \cline{2-17} 
\multicolumn{1}{|c|}{}                                            & \textit{\textbf{Precision}} & \multicolumn{1}{c|}{0.411}                   & \multicolumn{1}{c|}{0.435}                    & 0.432                        & \multicolumn{1}{c|}{0.411}                   & \multicolumn{1}{c|}{0.402}                    & 0.408                        & \multicolumn{1}{c|}{0.404}                   & \multicolumn{1}{c|}{0.379}                    & 0.42                         & \multicolumn{1}{c|}{0.333}                   & \multicolumn{1}{c|}{0.343}                    & 0.353                        & \multicolumn{1}{c|}{0.341}                   & \multicolumn{1}{c|}{0.371}                    & 0.346                        \\ \cline{2-17} 
\multicolumn{1}{|c|}{}                                            & \textit{\textbf{Recall}}    & \multicolumn{1}{c|}{0.494}                   & \multicolumn{1}{c|}{0.587}                    & 0.587                        & \multicolumn{1}{c|}{0.576}                   & \multicolumn{1}{c|}{0.57}                     & 0.593                        & \multicolumn{1}{c|}{0.831}                   & \multicolumn{1}{c|}{0.802}                    & 0.797                        & \multicolumn{1}{c|}{0.791}                   & \multicolumn{1}{c|}{0.86}                     & 0.849                        & \multicolumn{1}{c|}{0.733}                   & \multicolumn{1}{c|}{0.738}                    & 0.75                         \\ \hline
\multicolumn{1}{|c|}{\multirow{5}{*}{\textit{\textbf{Average}}}}  & \textit{\textbf{AUROC}}     & \multicolumn{1}{c|}{0.673}                   & \multicolumn{1}{c|}{\textbf{0.683}}           & \textbf{0.69}                & \multicolumn{1}{c|}{0.641}                   & \multicolumn{1}{c|}{0.634}                    & \textbf{0.643}               & \multicolumn{1}{c|}{0.71}                    & \multicolumn{1}{c|}{0.708}                    & \textbf{0.729}               & \multicolumn{1}{c|}{0.674}                   & \multicolumn{1}{c|}{\textbf{0.697}}           & \textbf{0.702}               & \multicolumn{1}{c|}{0.684}                   & \multicolumn{1}{c|}{\textbf{0.687}}           & \textbf{0.689}               \\ \cline{2-17} 
\multicolumn{1}{|c|}{}                                            & \textit{\textbf{F1-Score}}  & \multicolumn{1}{c|}{0.639}                   & \multicolumn{1}{c|}{\textbf{0.641}}           & \textbf{0.647}               & \multicolumn{1}{c|}{0.623}                   & \multicolumn{1}{c|}{\textbf{0.628}}           & \textbf{0.628}               & \multicolumn{1}{c|}{0.642}                   & \multicolumn{1}{c|}{0.628}                    & \textbf{0.644}               & \multicolumn{1}{c|}{0.637}                   & \multicolumn{1}{c|}{\textbf{0.641}}           & \textbf{0.655}               & \multicolumn{1}{c|}{0.645}                   & \multicolumn{1}{c|}{0.643}                    & 0.639                        \\ \cline{2-17} 
\multicolumn{1}{|c|}{}                                            & \textit{\textbf{MCC}}       & \multicolumn{1}{c|}{0.19}                    & \multicolumn{1}{c|}{\textbf{0.192}}           & \textbf{0.2}                 & \multicolumn{1}{c|}{0.244}                   & \multicolumn{1}{c|}{\textbf{0.255}}           & \textbf{0.255}               & \multicolumn{1}{c|}{0.278}                   & \multicolumn{1}{c|}{0.263}                    & \textbf{0.305}               & \multicolumn{1}{c|}{0.222}                   & \multicolumn{1}{c|}{0.214}                    & \textbf{0.227}               & \multicolumn{1}{c|}{0.211}                   & \multicolumn{1}{c|}{\textbf{0.211}}           & 0.2                          \\ \cline{2-17} 
\multicolumn{1}{|c|}{}                                            & \textit{\textbf{Precision}} & \multicolumn{1}{c|}{0.639}                   & \multicolumn{1}{c|}{0.637}                    & \textbf{0.642}               & \multicolumn{1}{c|}{0.659}                   & \multicolumn{1}{c|}{\textbf{0.662}}           & \textbf{0.662}               & \multicolumn{1}{c|}{0.666}                   & \multicolumn{1}{c|}{0.663}                    & \textbf{0.686}               & \multicolumn{1}{c|}{0.623}                   & \multicolumn{1}{c|}{0.603}                    & 0.618                        & \multicolumn{1}{c|}{0.618}                   & \multicolumn{1}{c|}{0.614}                    & 0.613                        \\ \cline{2-17} 
\multicolumn{1}{|c|}{}                                            & \textit{\textbf{Recall}}    & \multicolumn{1}{c|}{0.658}                   & \multicolumn{1}{c|}{\textbf{0.667}}           & \textbf{0.674}               & \multicolumn{1}{c|}{0.621}                   & \multicolumn{1}{c|}{\textbf{0.629}}           & \textbf{0.631}               & \multicolumn{1}{c|}{0.671}                   & \multicolumn{1}{c|}{0.658}                    & 0.658                        & \multicolumn{1}{c|}{0.697}                   & \multicolumn{1}{c|}{\textbf{0.723}}           & \textbf{0.733}               & \multicolumn{1}{c|}{0.699}                   & \multicolumn{1}{c|}{\textbf{0.7}}             & 0.689                        \\ \hline
\end{tabular}
    \end{adjustbox}
    \begin{minipage}{\textwidth}
\centering
        \scriptsize \textbf{Note:} Bold highlights improvements over P with respect to the corresponding evaluation metric.
    \end{minipage}
\end{table*}
\vspace{-2em}
\section{Results}
Table \ref{tab:CORRELATION TABLE} shows the correlation analysis to answer RQ1. It can easily be seen that co-change entropy is significantly correlated with the number of defects. For a better comparison, we also show the correlation results for change entropy with the post release defect count. It can easily be verified from the table that change entropy is better correlated with the defect count than the co-change graph entropy. However, both Pearson and Spearman correlation coefficients between co-change entropy and post-release defect counts are significant. It can also be seen that Pearson correlation coefficients are significantly more than Spearman coefficients, indicating the linear relationship between the independent variable (co-change graph entropy) and the dependent variable, i.e., post-release defect counts. The consistently low p-value (<0.05) across all projects indicates the statistical significance of our results. Based on this observation, we answer our first research question as follows:
\\\\
\fbox{\begin{minipage}{\dimexpr\columnwidth-2\fboxsep-2\fboxrule}
\textbf{Answer to RQ1:} \textit{Although the correlation between change entropy and post-release defects is higher than that of co-change entropy, co-change entropy remains significantly correlated with post-release defects, with a Pearson correlation coefficient exceeding 0.50 in some cases. Moreover, the higher Pearson correlation coefficient compared to Spearman’s suggests that the relationship between co-change graph entropy and post-release defects tends toward linearity.}
\end{minipage}}
\vspace{ 0.5em}
\par
\begin{figure*}[ht]
  \centering
  \begin{subfigure}{0.48\linewidth}
    \centering
    \includegraphics[width=\linewidth]{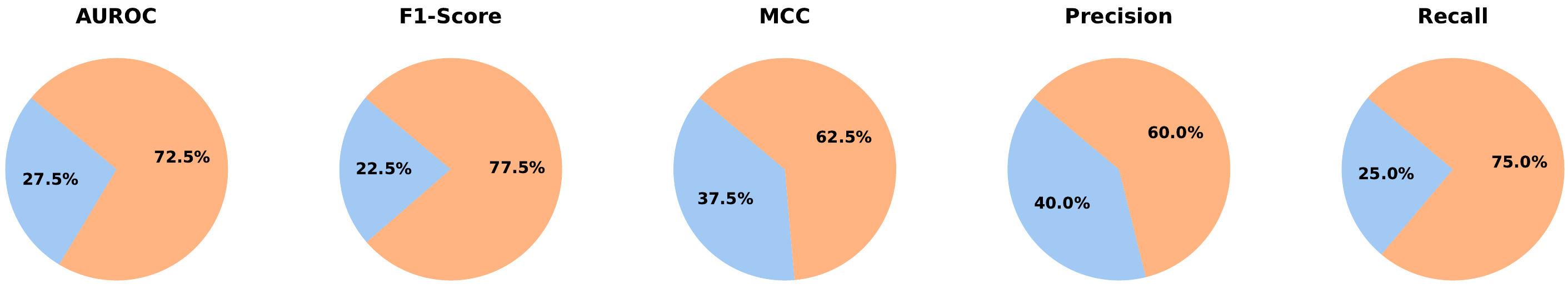}
    \caption{\small P+C (blue) vs.\ P+Co (orange)}
    \label{fig:changevscochange}
  \end{subfigure}
  \kern0.5em \vDotted{60pt} \kern0.5em
  \begin{subfigure}{0.48\linewidth}
    \centering
    \includegraphics[width=\linewidth]{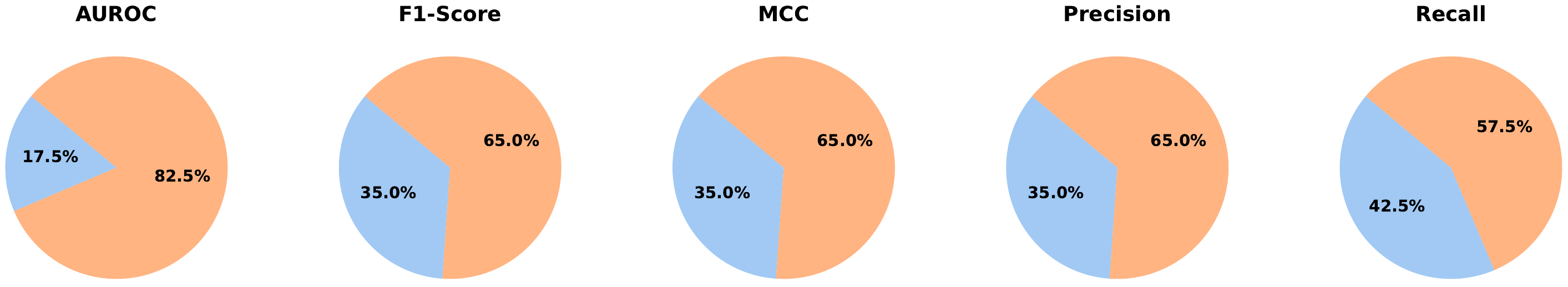}
    \caption{\small P+C (blue) vs.\ P+C+Co (orange)}
    \label{fig:changevspcco}
  \end{subfigure}

  \caption{Comparison of Predictive‑performance}
  \label{fig:combined_pie}
  \Description{Comparison of Predictive Performance}
\end{figure*}

 In defect classification, a single metric is rarely used. Instead, multiple metrics are combined to train machine learning classifiers, which then predict defect proneness in future releases. Thus, a metric’s high correlation with defect count alone is insufficient; it should complement existing metrics to enhance overall classification performance. To evaluate the effectiveness of our proposed co-change entropy in complementing other process metrics (as discussed in Section \ref{set up for RQ2}), we trained three classifier versions for each dataset-classifier pair: P+C, P+Co, and P+C+Co, and assessed their performance. Table \ref{tab:prediction} and Figure \ref{fig:combined_pie} summarize the results. Notably, replacing change entropy with co-change entropy in process metrics (P+Co) consistently outperforms the process metrics with change entropy (P+C). As shown in Figure \ref{fig:changevscochange}, across 40 dataset-classifier pairs (8 projects × 5 classifiers), co-change entropy improves AUROC in 72.5\% of the cases, F1-score in 77.5\%, MCC in 62.5\%, precision in 60\%, and recall in 75\%. 

\par Interestingly, although change entropy exhibits a stronger correlation with defect count than co-change entropy, classifier performance in defect prediction improves when co-change entropy replaces change entropy in the process metrics. This indicates that co-change entropy better complements other process metrics than change entropy. To analyze their combined effect, we compare classifier performance between P+C (process metrics with change entropy) and P+C+Co (process metrics with both change entropy and co-change entropy). The summary in Figure \ref{fig:changevspcco} shows that in 82.5\% of dataset-classifier pairs, AUROC is higher with P+C+Co than with P+C. A similar trend is observed across other performance measures, such as F1-score, precision, and MCC, where 65\% of the cases show an improvement with P+C+Co, while recall is improved in only 57.5\% of the cases. This demonstrates that co-change entropy complements change entropy in most cases, enhancing defect classification performance overall.


\par To statistically validate whether classifier performance differs significantly, we applied the non-parametric Friedman test followed by the post-hoc Nemenyi test, as recommended in prior research \cite{demvsar2006statistical}\cite{gong2019empirical}\cite{ lessmann2008benchmarking}. These tests were conducted across five performance metrics: AUROC, F1-score, precision, recall, and MCC. As shown in Table \ref{tab:friedman}, the Friedman test indicates that classifier performance differs significantly for AUROC, F1-score, precision, and MCC, but not for recall. The Nemenyi test revealed no significant difference between P+C and P+Co, but a statistically significant improvement with P+C+Co over P+C across all metrics except recall. These findings suggest that co-change entropy effectively complements process metrics, demonstrating its practical utility in defect prediction. In summary, the answer to our second research question is as follows:\\\\
\fbox{\begin{minipage}{\dimexpr\columnwidth-2\fboxsep-2\fboxrule}
\textbf{Answer to RQ2:} \textit{Substituting change entropy with co-change entropy in the process metrics set resulted in improved defect classification performance, with AUROC values increasing in 72.5\% of cases. However, this improvement was not statistically significant, failing the post-hoc Nemenyi test following the Friedman test. In contrast, combining change entropy and co-change entropy significantly enhanced defect classification performance, leading to AUROC improvements in 82.5\% of cases. This improvement was statistically significant, as confirmed by a successful post-hoc Nemenyi test after the Friedman test.}
\end{minipage}}
\begin{table}[h!]
    \centering
    \scriptsize
    \caption{Friedman and Post-hoc Nemenyi's Test}
    \label{tab:friedman}
    \begin{adjustbox}{max width=\textwidth}
    \begin{tabular}{|c|ccccc|}
    \hline
                                            & \multicolumn{1}{c|}{\textit{\textbf{AUROC}}} & \multicolumn{1}{c|}{\textit{\textbf{F1}}} & \multicolumn{1}{c|}{\textit{\textbf{MCC}}} & \multicolumn{1}{c|}{\textit{\textbf{Precision}}} & \textit{\textbf{Recall}} \\ \hline
    \textit{\textbf{Friedman P+C,P+Co,P+C+Co}}              & \multicolumn{1}{c|}{\checkmark}              & \multicolumn{1}{c|}{\checkmark}           & \multicolumn{1}{c|}{\checkmark}            & \multicolumn{1}{c|}{\checkmark}                 & x                        \\ \hline
    \textit{\textbf{Nemenyi P+C vs P+Co}}   & \multicolumn{1}{c|}{x}                       & \multicolumn{1}{c|}{x}                    & \multicolumn{1}{c|}{x}                     & \multicolumn{1}{c|}{x}                          & x                        \\ \hline
    \textit{\textbf{Nemenyi P+C vs P+C+Co}} & \multicolumn{1}{c|}{\checkmark}              & \multicolumn{1}{c|}{\checkmark}           & \multicolumn{1}{c|}{\checkmark}            & \multicolumn{1}{c|}{\checkmark}                 & x                        \\ \hline
    \end{tabular}
    \end{adjustbox}
    \begin{minipage}{0.45\textwidth}
\centering
        \scriptsize \textbf{Note:} `\checkmark' = p-value $< 0.05$ (The difference between the groups is statistically significant); \\`x' = p-value $> 0.05$ (The difference between the groups is not statistically significant).
    \end{minipage}
\end{table}
\section{Threats to validity}
\textbf{External Validity:} A key limitation of this study is its reliance on the SmartSHARK dataset \cite{trautsch2021msr}, which, although widely used, primarily consists of Java projects. This homogeneity limits the generalizability of our findings to software systems developed in other programming languages. While our project selection followed established criteria \cite{tantithamthavorn2018impact}, incorporating projects from more diverse datasets would strengthen the robustness and applicability of our results.
\par \textbf{Practical Validity:} Although the combined use of change and co-change entropy metrics resulted in statistically significant improvements in defect classification, the performance gain of 2–3\% remains relatively modest. This may limit the practical adoption of these metrics in real-world defect prediction models. Nevertheless, our findings highlight the potential of entropy-based graph measures and encourage further research into more advanced techniques that could yield greater predictive accuracy.
\section{Conclusion}
This study introduced co-change entropy, a novel metric that quantifies co-change dispersion in software development using co-change graphs. Our analysis revealed a significant correlation between co-change entropy and post-release defects, with Pearson correlation coefficients exceeding 0.50 in several projects. While replacing traditional change metrics with co-change metrics in defect classification led to performance improvements, these gains were not statistically significant. However, combining change and co-change entropy significantly enhanced prediction accuracy, as confirmed by statistical testing. These findings highlight the value of incorporating co-change entropy into process metrics to improve defect prediction. Our research is easily verifiable and reproducible, as we publish our code and all other artifacts necessary to reproduce our results in a publicly available replication kit\footnote{\url{https://github.com/hrishikeshethari/EASE2025}}
\par Software change history has been represented using more sophisticated graph constructs, such as hypergraphs and heterogeneous information networks, in previous studies. These advanced representations offer new opportunities for capturing software evolution patterns. For instance, heterogeneous graphs can model co-change dispersion at various types of node levels, providing a richer structural understanding. Extending this work by exploring heterogeneous graph entropy to capture such dispersions could be a promising direction for future research.
\vspace{-0.5em}
\bibliographystyle{ACM-Reference-Format}
\bibliography{References.bib}

\end{document}